# Detecting Urban Earthquakes with the San Fernando Valley Nodal Array and Machine Learning

Joses Omojola[1], and Patricia Persaud[1]

[1]Department of Geosciences, University of Arizona

Corresponding author: Joses Omojola (jomojo1@arizona.edu)

## Conflict of Interest Statement

The authors acknowledge that there are no conflicts of interest.

# Abstract

The San Fernando Valley, part of the Los Angeles metropolitan area, is a seismically active urban environment. Large-magnitude earthquakes, such as the 1994 $M_w$ 6.7 Northridge event that occurred on a blind fault beneath the valley, caused significant infrastructure damage in the region, underscoring the need for enhanced seismic monitoring to improve the identification of buried faults and hazard evaluation. Currently, the Southern California Earthquake Data Center operates four broadband instruments within the valley; however, the networks ability to capture small earthquakes beneath the region may be limited. To demonstrate how this data gap can be filled, we use recordings from the San Fernando Valley array comprised of 140 nodal instruments with interstation distances ranging from 0.3 to 2.5 km that recorded for one month. High anthropogenic noise levels in urbanized areas tend to conceal earthquake signals, therefore we applied a previously developed machine learning model finetuned on similar waveforms to detect events and pick seismic phases. In a two-step event association workflow, isolated phase picks were first culled, which eliminated false positive detections and reduced computational runtime. We located 62 events within a 209 km radius of our array, including 36 new events that were undetected by the regional network, with magnitudes ranging from $M_L$ 0.13 to 4. One event cluster reveals a previously unidentified (5.3 km by 4 km) blind fault zone located ~5 km beneath the southern part of the valley. Seismicity from this zone is rare in the regional catalog (<3 events per year), despite producing a $M_b$ 4.4 event in 2014. Our results highlight the benefits of detecting small-magnitude seismicity for hazard estimation. Temporary nodal arrays can identify critical gaps in regional monitoring and guide site selection for permanent stations. Additionally, our workflow can be applied to complement seismic monitoring in other urban settings.

# Introduction

The San Fernando Valley (SFV) in the greater Los Angeles area is a densely populated urbanized valley that is prone to damaging shaking from earthquakes (Figure 1; Hartzell et al., 1997; Krishnan et al., 2005; Vidale & Helmberger, 1988). The SFV basin initiated during an extensional period in the Miocene, following which the tectonic regime was altered to transpression during the evolution of the San Andreas fault (Crowell, 1979; Ingersoll, 2001; Levy et al., 2020). Due to the critical role that sedimentary basin structure plays in seismic wave amplification (e.g., Day et al., 2012; Graves & Wald, 2004; Liu & Heaton, 1984), the region is one of several in Southern California where substantial effort is being dedicated to advance knowledge of sedimentary basin structure and rock properties in order to improve preparedness for a future large earthquake on the San Andreas fault system (Ghose et al., 2023; Liu et al., 2018; Persaud et al., 2016; Villa et al., 2023; Yang & Clayton, 2024). In addition, two of the most devastating earthquakes in the Los Angeles area, the 1971 $M_w$ 6.7 San Fernando Valley and 1994 $M_w$ 6.7 Northridge earthquakes occurred on blind thrust faults beneath the SFV (Figure 1; Baldwin et al., 2000; Fuis et al., 2003) revealing large earthquake source regions as close as 25 km from downtown Los Angeles. Using information from the Northridge event, the United States Geological Survey (USGS) constructed new seismic hazard maps for Southern California that would account for the presence of blind faults (Updike et al., 1996). A study after the Northridge event showed that site amplification was largest for stations on softer soils and site amplification estimates show a better correlation with the detailed geology compared to a general geology classification (Bonilla et al., 1997; Catchings & Lee, 1996; Hartzell et al., 1997). The San Fernando and Northridge events have had far-reaching impacts on earthquake risk reduction (Hough et al., 2024) and underscore the need for improved understanding of both the seismic structure of sedimentary basins and hazardous faults in urban areas with high seismic risk.

Utilizing gravity and magnetic data, well logs, and 2D seismic reflection profiles, Langenheim et al. (2011) mapped the major faults across the SFV basin . The basin can be subdivided into the San Fernando basin in the south, and the Sylmar subbasin in the north that sits at a higher topographic elevation than the rest of the valley south of the Mission Hills fault (Levy et al., 2020). Basin depths from ambient noise tomography and gravity modeling show the detailed structure of the SFV basin with its deepest part occur in the Sylmar subbasin, and strong velocity contrasts characterizing the major faults (Juárez-Zúñiga & Persaud, 2025). The southern edge of the basin hosts normal faults from the extensional episode, while the northern part has thrust faults from the transpressional episodes (Levy et al., 2020). The two major thrust faults beneath the SFV are the Oak Ridge fault zone (ORFZ) and the San Fernando fault which dip to the south and north respectively, and the ORFZ is overthrust by the Santa Susana fault (Figure 1; Jennings et al., 2010). Movement along the thrust faults is uplifting the Sylmar basin at a faster rate than the SFV, and causing seismicity beneath the basin (Langenheim et al., 2011; Levy et al., 2020).

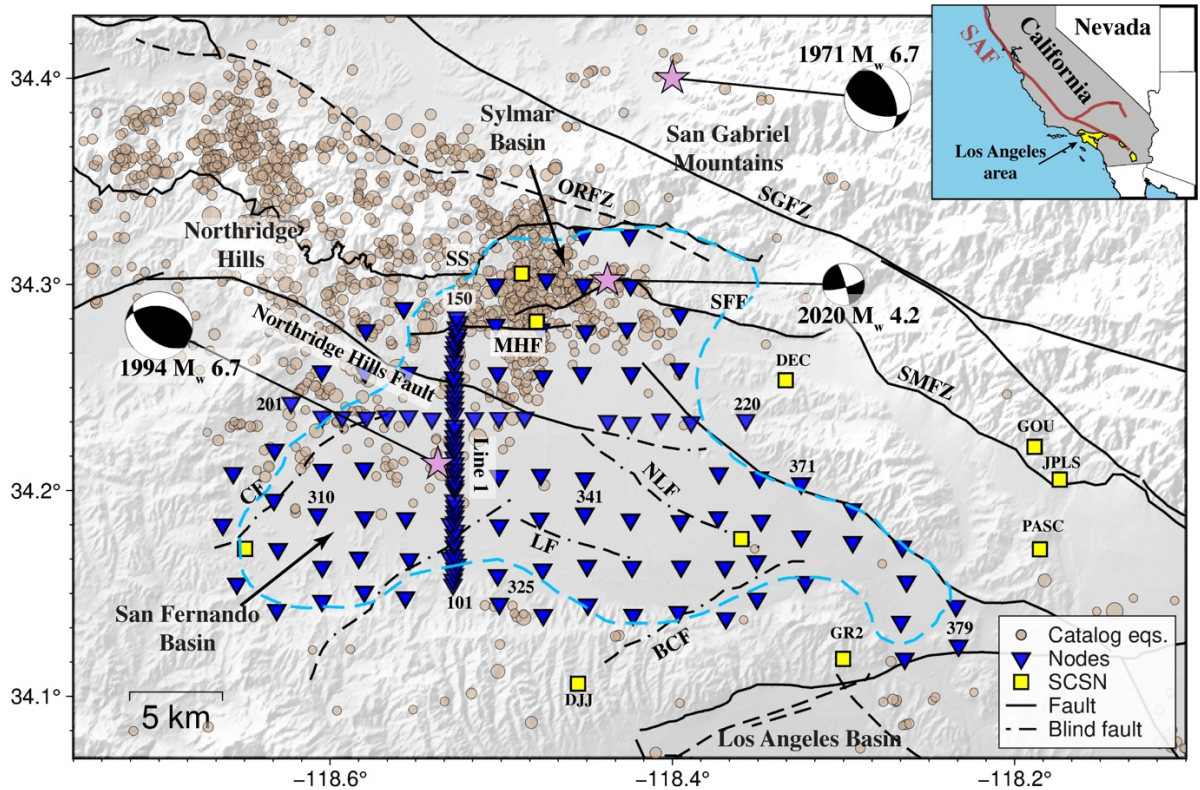

Figure 1. Topographic map of the study area showing the San Fernando Valley nodal array and the Southern California Seismic Network (SCSN) stations used in this study with selected stations labeled. Black lines are faults from the SCEC Community Fault Model (CFM) 6.1 (Marshall et al., 2023). Focal mechanisms of the 1971 San Fernando and 1994 Northridge earthquakes are shown (USGS, 2025). Brown circles are the $M_w>2.5$ earthquakes from the Hauksson et al. (2012) catalog from 1981 to 2018. Yellow fill in the inset map represents the outline of the major sedimentary basins in Southern California. Major faults and fault zones are the MH-Mission Hills fault, SAF-San Andreas fault, SF-San Fernando fault, SS-Santa Susana fault, ORFZ-Oak Ridge Fault Zone or Northridge thrust, SGFZ-San Gabriel Fault Zone, and SMFZ-Sierra Madre Fault Zone. Additional faults from Juárez-Zúñiga and Persaud (2025) are BCF-Benedict Canyon fault, CF- Chatsworth fault, LF-Leadwell fault, and NLF-North Leadwell fault.

The seismicity forms diffuse clusters around the surface traces of the major faults in and around the northern part of the basin (Figure 1). Regional catalogs show the majority of these events are part of the aftershock sequence of the 1994 Northridge event (Hauksson et al., 2012).

While the seismic hazard north of the basin is better understood, less is known about the seismicity and structural characterization in the southern part of the valley.

Local seismicity is useful for improving our understanding of earthquake statistics and seismic hazards, however small magnitude events (M<~1.8) can be routinely missed by regional networks in sparsely instrumented areas (Hutton et al., 2010). In a study in metropolitan Los Angeles, Walls et al. (1998) suggest that small-magnitude seismicity along short fault segments is more likely to stimulate seismic moment release compared to infrequent large magnitude events along major faults. The short recurrence intervals of small-medium magnitude events also make them useful in other urban seismic hazard studies such as evaluating site effects (Islam et al., 2025), mapping seismic scattering and absorption (Nardoni & Persaud, 2024), and developing high-resolution velocity models for improved ground motion simulation studies. Within the SFV, the Southern California Earthquake Data Center (SCEDC) currently operates four broadband instruments under the CI network (Figure 1) for monitoring seismicity (Caltech & USGS, 1926). Additional broadband stations can be costly to install and maintain, and have a large footprint, however, temporary dense nodal arrays can be employed to supplement broadband instruments, and address monitoring gaps. The advancement of nodal seismometers which are less expensive and easier to install have made nodal experiments and large-N arrays (>100 instruments) more commonplace. Their autonomous operation makes nodes ideal for short-term experiments such as monitoring aftershocks and complementing regional networks (Catchings et al., 2020; Farrell et al., 2018; Lythgoe et al., 2021).

There is general consensus that detecting small-magnitude events is useful, but the task of detecting these events in urban areas is limited by the high anthropogenic noise levels. High noise levels in urban environments such as the SFV can mask seismic signals complicating event detection and phase picking (Castillo et al., 2024; Fiori et al., 2023; Vassallo et al., 2024). Although dense nodal deployments can improve the chances of event detection, relatively large datasets can further complicate the manual analysis of small magnitude events (Di Luccio et al., 2021; Liu et al., 2020; Yang & Clayton, 2023). Recent advancements in machine learning (ML) models for automated event detection and phase picking have demonstrated their efficiency at processing large datasets (Mousavi et al., 2020; Zhu & Beroza, 2019). Compared to other automated methods like template-matching and short-term average/long-term average, ML techniques provide an optimal tradeoff between processing speed and detection accuracy (Kubo et al., 2024; Münchmeyer et al., 2022). They generalize well to new datasets, and can be readily finetuned to improve prediction accuracy on complex datasets (Woollam et al., 2022). By leveraging these advancements, we can improve the detectability of small magnitude events in densely populated urban environments and shed light on the limitations of such detection efforts. In October to November 2023, an array of three-component SmartSolo nodal seismometers was installed across the San Fernando Valley to test the feasibility of addressing seismicity monitoring gaps and to develop a high-resolution 3D seismic velocity model of the sedimentary basin. Here, we apply a pretrained ML model for automated processing of the dataset, with the goal of evaluating the limitations of detecting small-magnitude events in an urban setting and improving seismic hazard characterization beneath the San Fernando Valley.

## Dataset

The SFV array was installed by a team of volunteers that included individuals from universities in Arizona, California, Louisiana, Texas, and Mexico, as well as non-academic volunteers. The teams discussed the objectives of the project with residents and business owners and obtained permission to install the instruments on their properties, or installed them in public places (Persaud et al., 2024). The nodes were buried in ~20 cm deep holes as far away as possible from roads to reduce traffic noise. In total, 140 three-component SmartSolo IGU-16HR 3C nodal seismometers were installed across the San Fernando Valley over an area of ~520 km$^2$. The instruments were split into 3 subarrays (Figure 1), with the first subarray being a dense line of 49 stations installed along a 15 km profile spaced ~300 m apart. This subarray was installed near the epicenter of the 1994 Northridge earthquake to improve seismic characterization across the Northridge Hills and Northridge thrust faults. A second 15-node linear subarray was installed along a W-E profile, and a grid of 76 stations with an interstation distance of ~2500 m was installed across the rest of the valley. The nodes recorded at 500 Hz from 21 October to 22 November 2023, and the recordings were manually downloaded from the nodes after they were retrieved. We also analyzed data from 10 broadband stations deployed within and around the SFV. These stations are part of the CI network and continuously record data at 100 Hz. Instrument response files for the nodal and broadband stations were obtained from the Earthscope Seismological Facility for the Advancement of Geoscience (SAGE) Data Management Center, and the waveforms were converted to displacement. The dataset is available at the Earthscope SAGE Data Management Center (Persaud, 2023). We also utilized several existing datasets to provide context for our analysis and evaluate potential relationships with known fault structures. We obtained catalogs of relocated seismicity (1981-2018) and focal mechanisms in Southern California from the SCEDC special datasets (Hauksson et al., 2012; Yang et al., 2012). The aftershock catalog of the 1971 San Fernando earthquake was accessed through the SCEDC (Mori et al., 1995) and 3D fault plane representations were obtained from the SCEC Community Fault Model (CFM) 6.1 (Marshall et al., 2023). Fault names correspond to those in the SCEC CFM 6.1.

## Methodology

### Event Detection

We began our analysis by creating a catalog of seismic phases associated with events recorded by the nodal array. ML models are reliable for automating earthquake phase detection and they generalize well to unseen datasets (Kubo et al., 2024). Compared to traditional methods like short-term average / long-term average (STA/LTA) and template matching cross-correlation catalogs, ML models provide an optimal tradeoff between phase detection accuracy, and computation runtimes (Kubo et al., 2024). For phase detection, we used a hybrid autoencoder machine learning model developed to detect small-magnitude seismicity in high-noise industrial settings, hereafter referred to as microearthquake-Unet (MUnet) (Omojola & Persaud, 2024). The MUnet model was developed to achieve high-quality phase detection for microearthquakes which are important for accurate event locations while minimizing false positives caused by non-earthquake signals. It was first trained on small magnitude earthquakes ($M_w < 3$) then fine-tuned on microearthquakes ($M_L < 2$) recorded by nodal arrays in an industrial area. MUnet has been tested and compared to other ML models and is well suited for the SFV dataset. The model processes three-component waveforms downsampled to 100 Hz, and outputs phase pick probabilities for both P-wave and S-wave arrivals. In this study, we cut our waveforms into 6-s long windows with 25% overlap. A pick probability threshold of 0.4 was used for detecting phase picks.

### Event Association and Location

Following phase identification, the P- and S-phase picks were processed in an event association workflow. Pick association is a process of merging phase arrivals from multiple stations to create a catalog of events. Simple association techniques like grouping the number of picks within a time window (e.g. 2 s) generates subpar association results because of the moveout time across the entire array, especially for S-waves that are slowed by the underlying sediments. More robust association techniques that utilize velocity models are more accurate for phase association but trade off computational speed for accuracy. As the number of picks per event and the total number of events increase, these association algorithms can stall for several days without recovering or develop segmentation faults (Münchmeyer, 2024). Preliminary tests with the Gaussian Mixture Model Association (GaMMA) algorithm did not provide encouraging runtime results for known Southern California Earthquake Data Center (SCEDC) catalog events; therefore we did not test it on the entire dataset (Zhu et al., 2022). In this study we tested both the Rapid Earthquake Association and Location (REAL) and Pyocto algorithms for event association (Münchmeyer, 2024; Zhang et al., 2019) . REAL and Pyocto produced similar results for synthetic tests that we executed using the SFV array station configuration. Pyocto's computational runtime was faster than REAL, so we opted to apply it to our full dataset.

The high volume of picks across the 140 stations (~1400 phase picks per station per day) required a computation time of over 24 h with the Pyocto algorithm. To improve the runtimes, we developed an association preprocessing algorithm (Figure 2a-b). Our algorithm first sorts all the phase pick arrival times and uses a cached table of interstation distances to estimate interstation apparent seismic velocities based on consecutive travel-times. Minimum apparent seismic velocities ($V_P$ - 3.5 km/s and $V_S$ - 2 km/s) were then used to eliminate slow uncorrelated picks, and a minimum of 12-consecutive picks were further required for event identification. The minimum apparent velocity cutoff enables identification of pick clusters with interstation travel-times that fit reasonable body wave speeds for rays traveling through both sedimentary and crystalline rocks (Christensen and Mooney, 1995). This eliminates isolated or random picks from cultural noise sources and significantly reduces the number of input picks to the association algorithm (Figure 2b). After preprocessing, the association computational time reduced from >24 h to < 8 h. We required a minimum number of 12 associated phase picks to identify an event, using a 1D velocity model of southern California (Hadley & Kanamori, 1977). Increasing the minimum number of associated picks can reduce computational runtimes even further (<1 h) at the expense of skipping small microearthquakes.

    The association algorithm provides preliminary locations, however, we use the HYPOINVERSE and HYPODD programs to improve the event locations (Klein, 2002; Waldhauser, 2001). Python wrappers were developed around binary executables for both programs to enable easier location and relocation of the events in a single environment. We refined our initial event locations from HYPOINVERSE using the HYPODD double-difference relocation method. This helped to reduce our travel-time residuals to below 0.06 s and improved our catalog locations. Distant larger magnitude regional events (>30 km from our array) that were recorded on our array did not form sufficient relocation clusters; therefore, for these events, we used initial location estimates from HYPOINVERSE with travel-time residuals (<1 s) in our catalog.

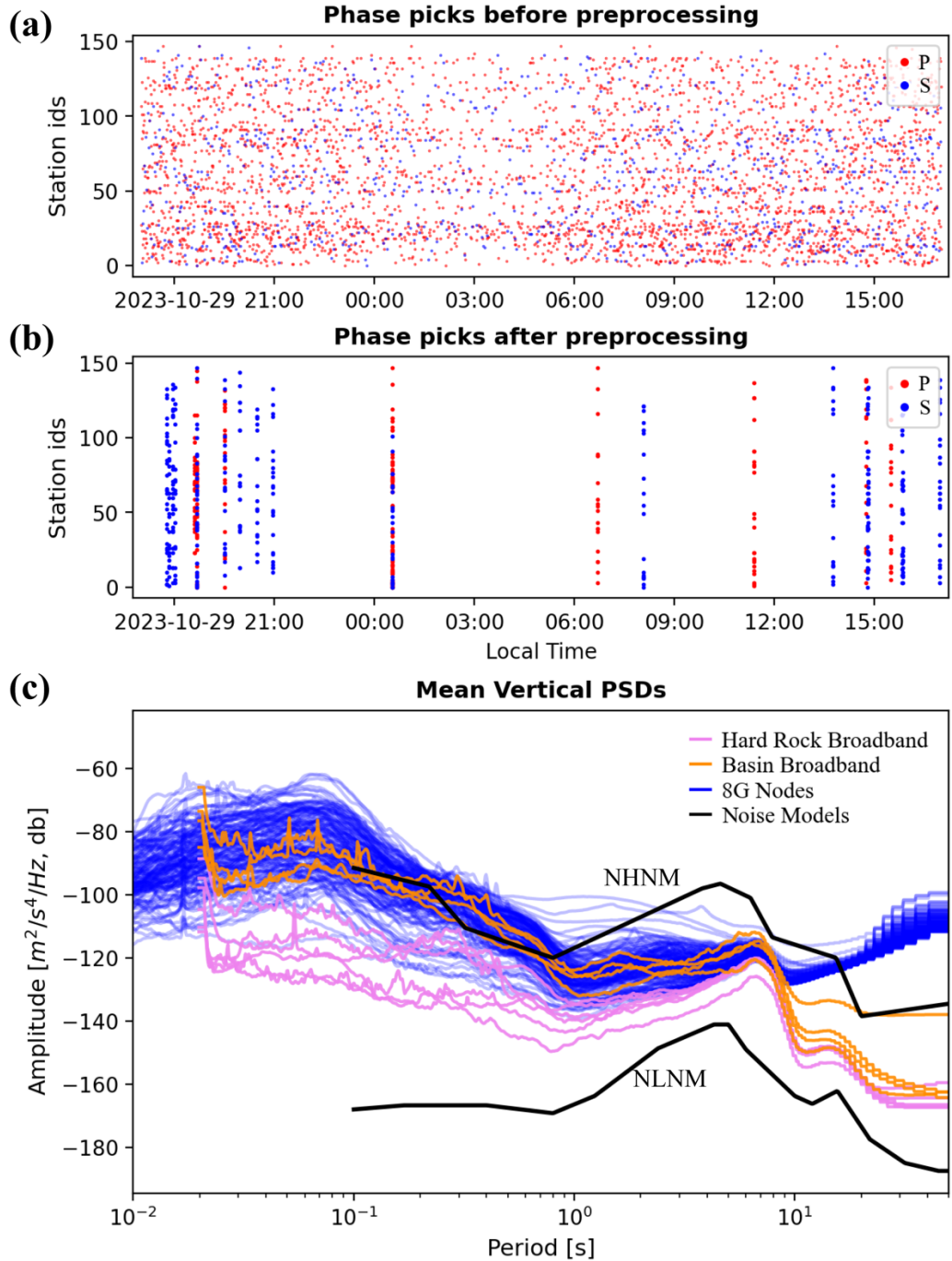

Figure 2. Preprocessing of phase picks for event association and comparison of noise levels of the node and broadband stations. (a) Raw picks before preprocessing from October 30, 2023 (UTC) with the x-axis shown in local time. (b) Preprocessed picks after removing isolated detections. Fewer picks speed up the association process. Similar plots for November 05, 2023 (UTC) are shown in Figure S5. (c) Mean noise

levels for vertical components of nodal and broadband stations for the full recording period. Relatively quiet broadband instruments are outside the SFV basin, and collocated nodal and broadband instruments within the basin have similar noise levels for periods below 10 s. NHNM: New High Noise Model, NLNM: New Low Noise Model (Peterson, 1993).

Event Magnitudes and Detectability

Local magnitudes, $M_L$ of relocated events were estimated using a regression equation (1) that is dependent on the maximum amplitude of the P-wave, A in mm, and the epicentral distance, Δ for each event-station pair (Richter, 1935).

$$M_L = \log_{10}A + a\log_{10}\Delta + b \quad\quad\quad 1$$

Eighteen events from our relocated catalog, with known magnitudes from the SCEDC catalog were used to calibrate local magnitude values. Optimal estimates of the empirical values a and b were obtained by linear regression, and our best fit results were *a* = 1.7175 and *b* = 6.1777. Magnitudes of local events detected by our array that are not in the SCEDC catalog were estimated using (1), and the median value across all recording stations was used to calculate the catalog magnitudes.

We estimated focal mechanisms for local events using the HASH method (Hardebeck & Shearer, 2003). Because waveform moment tensor inversions are not always suitable for high-frequency microearthquakes, we utilized P-wave first motion polarities and a 1D velocity model to calculate fault plane orientations with uncertainty bounds. Our approach required a minimum of eight stations with alternating polarity directions to obtain quality focal mechanism solutions.

Transient influences on urban recordings due to local site conditions while negligible for relatively long-period regional and teleseismic events (Liu et al., 2018), play a large role in determining the detectability of local events by our array. Event detectability at individual stations depends on different factors including but not limited to local geology, nearby noise sources, station coupling, event magnitude, and hypocentral distance (Inbal et al., 2024). To evaluate the detectability of our array, we began by assessing the background noise levels using the probabilistic power spectral density (PPSD) of individual stations. The PPSDs were calculated by slicing continuous waveforms into 30 minute windows with 50% overlap, and the power spectrum was estimated using the routines of McNamara and Buland (2004) in obspy (Krischer et al., 2017). The mean PPSDs provide insights into local noise levels for different instruments (Figure 2c), and the background seismic noise at 15-minute intervals was converted to displacement using the routines of Lecocq et al. (2020), to estimate noise levels at specific times for detected events.

Given that the SFV array recorded both local and regional events, using only epicentral distance or magnitude cutoffs to determine detectability limits can be inaccurate, as both distant large-magnitude and nearby small-magnitude events may still be detected. To assess the seismic signal at individual stations, we rely on ground motion prediction equations (GMPE), because they are useful for estimating ground shaking intensity while accounting for earthquake properties like magnitude, distance to site, and spectral period (Boore & Atkinson, 2008). The peak amplitude of displacement (*Pd*), which is an example of a GMPE, provides a regression-based estimate of ground motion, and is derived from the empirical relationships between magnitude and hypocentral distance (Wu & Zhao, 2006). Using 684 vertical-component records, Wu and Zhao (2006) derived regression estimates of *Pd* empirical coefficients for $M_w < 7$ events in Southern California as

$$log(P_d) = -3.463 + 0.729M - 1.374log(R) \pm 0.305 \qquad 2$$

where M is the event magnitude, R is the hypocentral distance, and the unit of $P_d$ is cm. The $P_d$ of relocated events was converted from displacement, $disp_{cm}$ to decibels, $amp_{db}$ at individual stations using 3.

$$amp_{db} = 20 \log_{10}(disp_{cm}) \qquad 3$$

The detectability threshold is calculated by subtracting the empirical signal amplitude from the nearest background noise amplitude at frequencies between 1-20 Hz. Examples of the detectability threshold are shown in Figure S1. When the signal amplitude exceeds the background noise amplitudes, the event is classified as being detectable at that station. The integration of event magnitudes, hypocentral distances, and noise levels at individual stations allows estimation of dynamic detectability thresholds. These thresholds are adaptive to temporal changes in noise levels, which is expected for an urban area like the San Fernando Valley and provides a robust estimate of event detectability at each station.

## Results

We detected and located 62 events, within a 209 km radius of the SFV array, during the month-long recording period (Dataset S2). The median signal-to-noise ratio (SNR) for daytime events across all stations was 0.34 compared to 4.2 for nighttime events, and many of our catalog events (61%) were detected at night between 8 pm to 5 am Los Angeles time. Thirty-six events were relocated, with 17 of them occurring near faults within 10 km of our array. Average location errors for our relocated catalog were 108 m and 152 m in the horizontal and vertical directions with average travel time residuals of ~10 ms. Location errors were relatively higher for the full catalog which includes events outside our station's coverage. Our relocated events are shown in Figure 3, and our full catalog is shown in Figure S2. During the same period, the SCEDC catalog lists ten events within 10 km of our array, and we recorded nine of those events. The tenth undetected M1.48 event is considered a spurious event, because its location is ~3 km from our array, but nearby nodal and broadband stations show no evident waveforms for this event (Figure 3, S3). We also detected 15 additional events that were undetected by the SCEDC but could not be relocated reliably within the SFV (Figure S2). These events were generally smaller, recorded on fewer stations (< 24) in our array, and had higher travel-time residual errors. The station density of our nodal array provides improved constraints on our computed locations within the valley, giving us better resolved hypocenters and depth estimates compared to the SCEDC catalog (Figure 4).

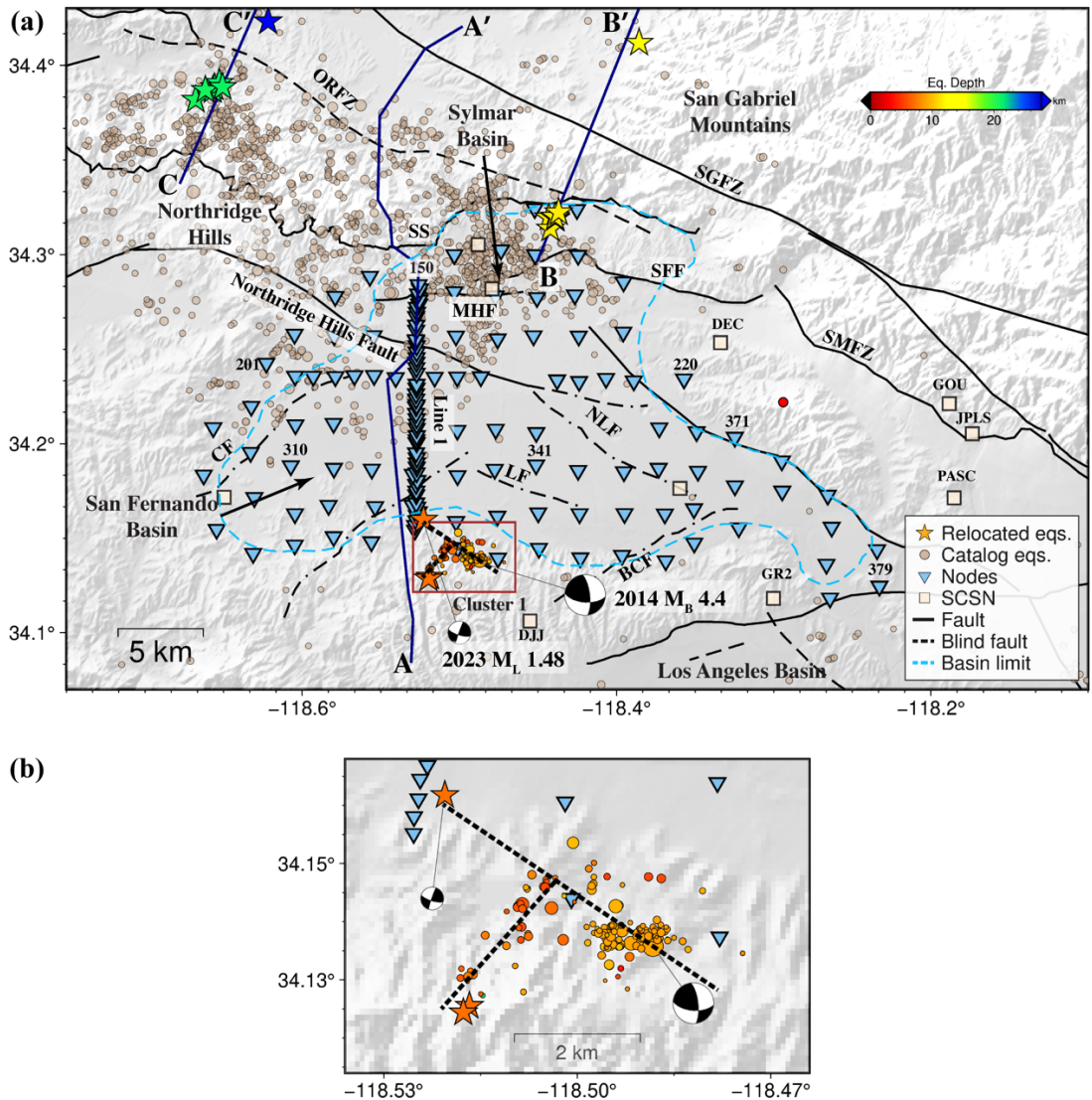

Figure 3. Map of relocated local events within 10 km of installed stations colored by hypocentral depth. (a) Relocated events are shown as stars. The locations of the A-A', B-B', and C-C' cross-sections in Figure 5 are shown with solid blue lines. Fault names and station are the same as Figure 1. The red circle near node station 371 marks the location of the spurious SCEDC catalog event with waveforms shown in Figure S3. Cluster 1 with three relocated events is highlighted with the brown rectangle. (b) A zoomed-in image of the cluster 1 events. A representative fault plane that aligns with the M> 4.4 aftershock seismicity from Hauksson et al. (2012) is shown with the dashed black line.

Our relocated event magnitudes around the SFV range between $M_L$ 0.13 to 2.04, with a median magnitude of $M_L$ 1.27. Event depths range between 6 - 25 km, forming mini-clusters along major thrust faults like the Oak-Ridge fault zone and the Santa Susana faults north of the SFV (Figure 3, 5a-c). Three of these events were located south of our array, in an area without any mapped active faults and we designate these events as cluster 1 (Figure 3). The events are located within proximity of prior seismicity in the Hauksson et al. (2012) catalog and a concealed Pre-Quaternary fault in the California Geologic Survey records (Hauksson et al., 2012; Jennings et al., 2010). Map and cross-section estimates of the fault zone indicate potential intersecting fault branches with a dimension of ~5.3 km by 4 km. Our focal mechanism solution for the largest magnitude ($M_L$ 1.48) event is well constrained with a HASH quality "B", and has a strike, dip, and rake of 108°, 73°, and 176°, respectively (Figure 4). The event is located 7.04 km beneath the basin (Figure 5a) and can be interpreted as a dominantly strike-slip fault with a minor thrust component. The remaining two events had magnitudes of $M_L$ 0.13 and 0.44 respectively, and reliable focal mechanisms could not be computed for these events because we had insufficient opposite first polarity measurements. The remaining 19 events in our relocated catalog were a combination of local and regional events from outside of the SFV area. Epicentral distances ranged between 29 - 209 km, and local magnitudes for these events ranged between $M_L$ 0.77 to 4.0. The SCEDC catalog contained 12 of these 19 events, and all events were detected by at least 6 stations in our study.

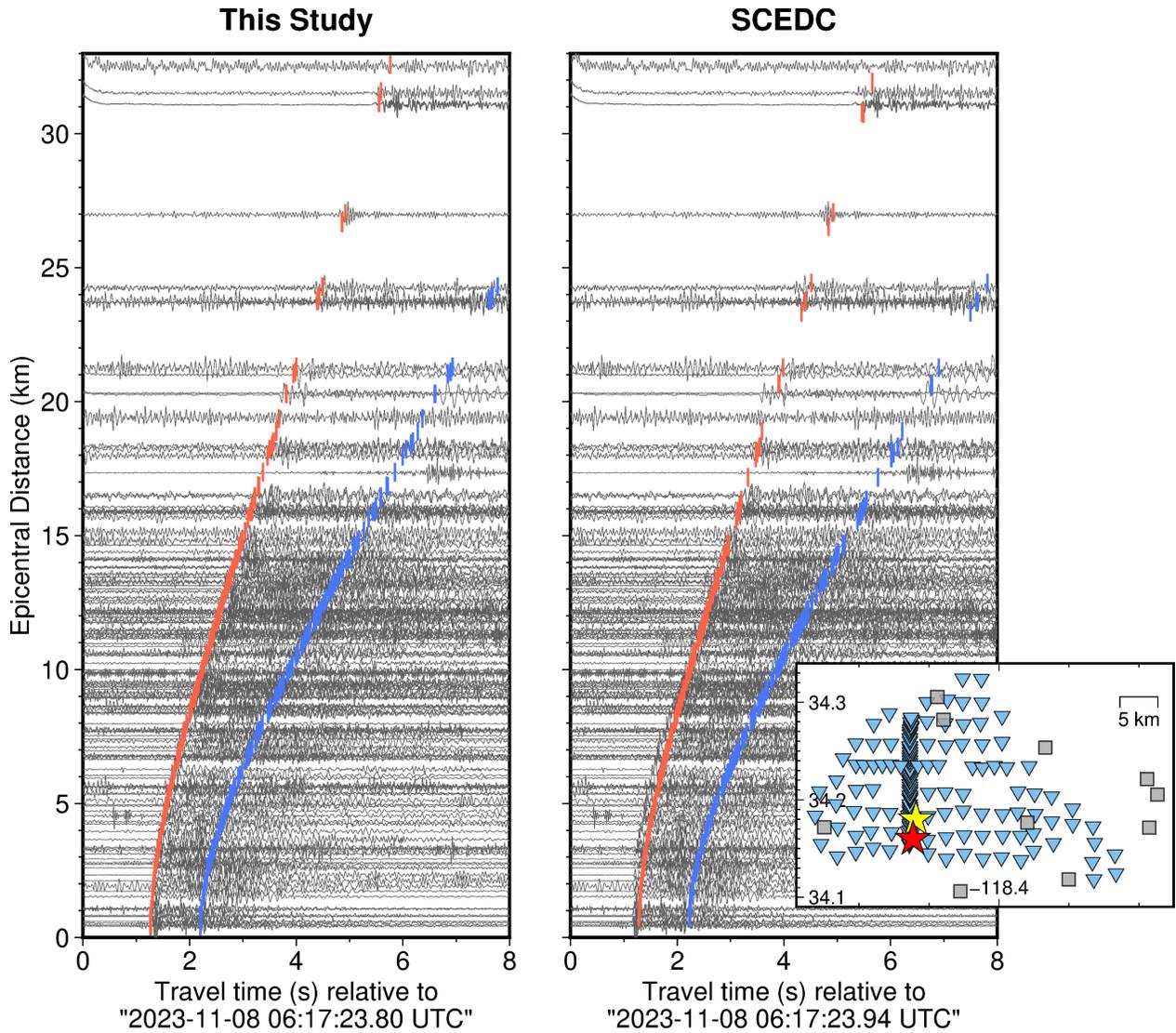

Figure 4. Comparison of predicted P- and S-wave arrival times for two different event locations for the $M_L$ 1.48 event shown in Figure 3. P- and S-wave arrivals in red and blue, respectively, were calculated using a 1D velocity model (Hadley & Kanamori, 1977) to compare event locations from our study (left) and the SCEDC catalog (right) shown in Figure 6b. Waveforms are sorted by epicentral distance. The SCEDC hypocenter gives a poor match and appears later than observed arrivals for stations <10 km from the epicenter. Different start times were used to crop the waveforms to ensure consistency with both catalogs. (inset) The epicentral locations from the SCEDC and this study for the $M_L$ 1.48 event shown in Figure 3. Nodal and broadband stations are shown as blue triangles and grey squares, respectively.

We reviewed the SCEDC catalog seismicity in the SFV from the last decade (2010-2023), to evaluate the earthquake statistics around the SFV. We selected a time period where no major aftershocks or magnitude modifications were implemented to reduce the risk of masking natural changes in seismicity (Tormann et al., 2010). A total of 2510 events were included in our search criteria, and the magnitude of completeness ($M_C$) of the catalog is 2.5. This indicates that the regional network is capable of reliably detecting M2.5 events and greater. The b-value was calculated using the Gutenberg-Ritcher relationship which is a function of the relative proportion of small to large earthquakes (Geffers et al., 2022). The estimated b-value of 0.85 is on par with regional studies of seismicity in southern California (Hutton et al., 2010; Tormann et al., 2010). Although the time period of our catalog is too restricted to allow a meaningful comparison to $M_C$, the magnitude frequency distribution of our catalog includes a higher proportion of smaller magnitude events M<0.5 (Figure 6). This suggests that the smaller magnitude events below 0.5 may be routinely undetected by the regional array, and additional permanent stations can help lower the overall magnitude of completeness within the SFV.

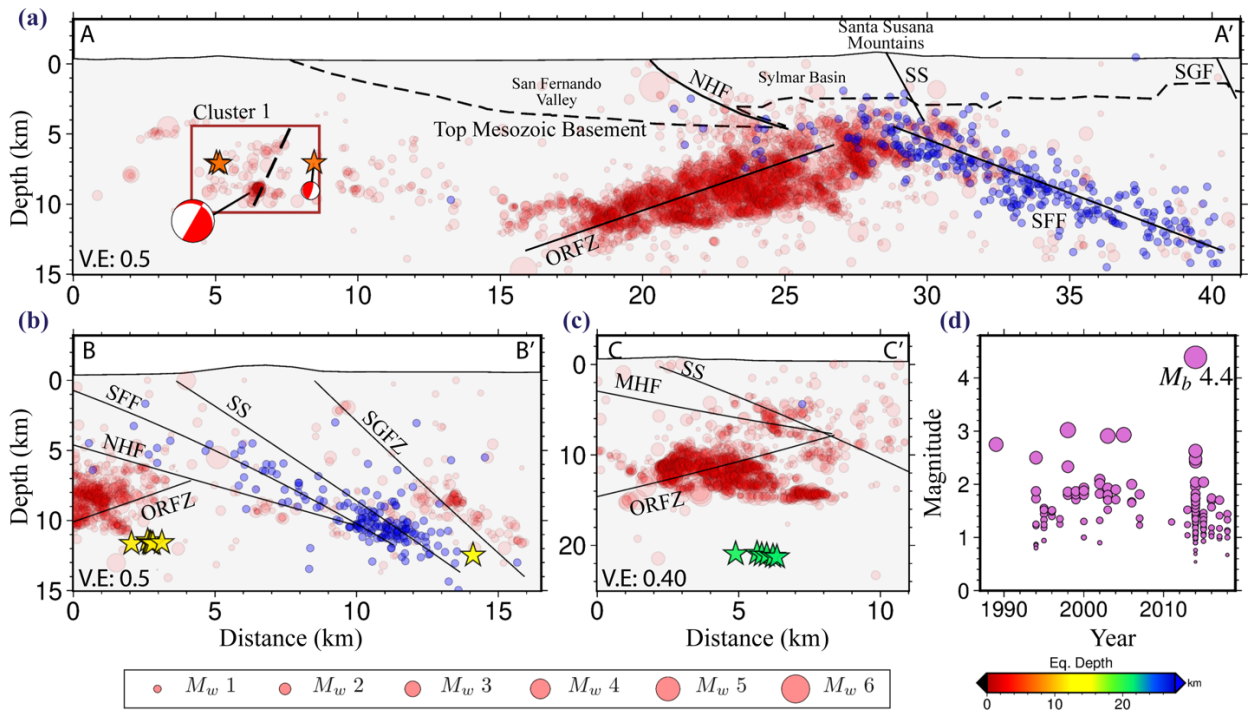

Figure 5. Cross-sections showing our relocated events with stars colored by hypocentral depth, the background seismicity with red circles (Hauksson et al., 2012), and aftershocks from the 1971 San Fernando earthquake with blue circles (Mori et al., 1995). Profile locations are shown in Figure 3. (a) Projected seismicity within 4 km of profile A-A'. The ORFZ and SFF faults converge beneath the San Fernando Valley. Basin depth and known faults are modified after (Langenheim et al., 2011). The background seismicity shows the interpreted dip of the cluster 1 fault zone. Focal mechanisms for the $M_L$ 1.48 and $M_b$ 4.4 events indicate an oblique strike-slip fault, with a dip ranging between 71-72°, and minor reverse component. (b) Projected background seismicity within 2 km of profile B-B'. Projected fault locations from the SCEC CFM 6.1 are shown as solid lines. Relocated events are linked to seismicity along the SGFZ and ORFZ. (c) Projected seismicity and faults along profile C-C'. Our relocated events are located below the major event clusters. (d) Magnitude versus time distribution of events in the Cluster 1 fault zone from Hauksson et al. (2012) showing quasi-continuous and ongoing activity.

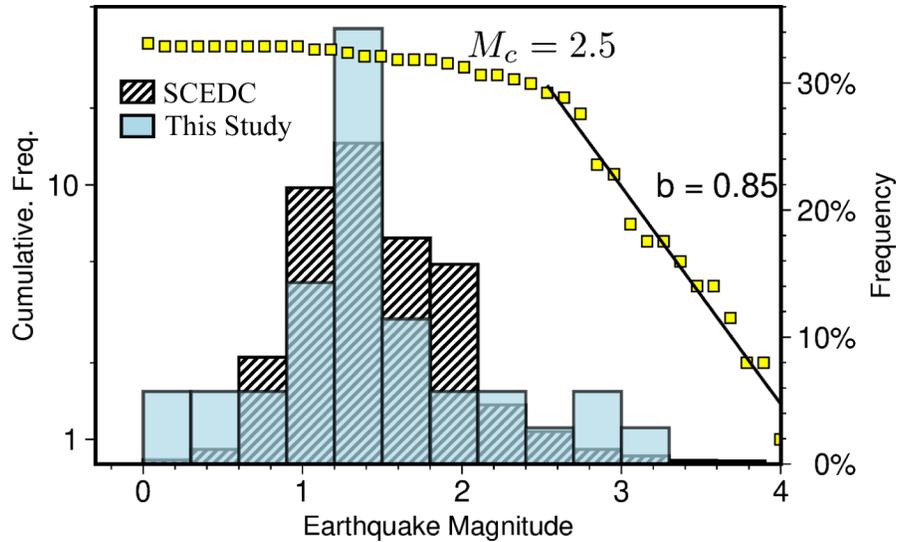

Figure 6. Magnitude frequency distribution for our catalog compared to the SCEDC catalog of events around SFV (2010-2023). The blue histogram includes all 36 relocated events detected in this study.

## Discussion

A major challenge in urban seismic monitoring is the high background noise levels caused by anthropogenic activities such as transportation, construction, and industrial operations (Bao et al., 2024; Inbal et al., 2018). Temporal differences in noise levels can complicate earthquake detection reinforcing a need to estimate event detectability which can be used to modify event detection strategies. For example, the $M_L$ 0.13 event in cluster 1 occurred at night and was recorded at a maximum epicentral distance of 18 km, while the $M_L$ 0.44 event which occurred during the day, was recorded at a maximum epicentral distance of 9 km. In the SFV, higher daytime noise levels impact event detection as evidenced by the lower median signal-to-noise ratio (SNR) we observed for daytime events (0.34) compared to nighttime events (4.2). Similar issues have been encountered in other urban studies such as in Auckland, New Zealand, London, UK, and the Long Beach and Seal Beach areas of Southern California, (Boese et al., 2015; Green et al., 2016; Yang & Clayton, 2023). To address noise contamination, some authors focus on analyzing mainly night-time events, deploying borehole instruments, or employing ML denoisers (Boese et al., 2015; Yang et al., 2022; Yang & Clayton, 2023). But borehole instruments are expensive to install and employing trace-by-trace ML denoisers can attenuate coherent signals in waveforms, reducing the overall quality of phase picks and detected events (Yang & Clayton, 2023).

In the SFV, we adopted the MUnet model that was finetuned on noisy event waveforms for event detection and phase-picking. This allows detection of small-magnitude events with the tradeoff of a high number of phase detections (Figure 2a). We designed our preprocessing algorithm to optimize the computational efficiency of event association, eliminate spurious phase detections, and reduce the association runtime without sacrificing accuracy. Using our initial catalog of detected events and station noise estimates from PPSD calculations, we were able to estimate event detectability for the $M_L$ 1.48 event at local stations which compares well to the observed detectability (Figure S1e-f). When applied in a permanent network scenario, computed detection thresholds can be used to dynamically adjust the ML picker probabilities at different times of the day when higher noise levels are expected. This can help reduce the overall false positive events from daytime cultural noise sources and improve the event detection quality without compromising catalog completeness by omitting real daytime events or compromising coherent signals with denoising.

The magnitude of completeness of a catalog provides insights into the network's detection capability. A lower magnitude of completeness can indicate a higher detection capability for small magnitude events (Schorlemmer & Woessner, 2008). Despite the short duration of our deployment, the high station density resulted in our array detecting a large proportion of small-magnitude events within the SFV, across multiple stations (Figure S2). This is expected to increase the ratio of events with magnitudes < 1.5 over time, contributing to an overall lower $M_C$. For some events outside of our array coverage, our relocation workflow gave deeper locations than seismicity in the SCEDC catalog. This may be due to the sparse azimuthal coverage in the phase arrivals used to locate these events (C-C' in Figure 5). Within the SFV, increased P- and S- phase picks from our dense network potentially improves constraints on our locations including event depths, which results in a better fit of predicted arrival times for near epicentral stations.

The importance of our results extends beyond event detection and association. Prior studies of fault activity highlight detailed shallow crustal structures (<5 km) using a combination of well logs, geophysical data, and forward structural modeling (Langenheim et al., 2011; Levy et al., 2020). While these structures are generally important, the large magnitude earthquakes in the SFV occur on deeper faults >10 km, where the geometry and inter-fault relationships are not as clearly defined. Fault traces from the USGS and 3D fault planes from SCEC CFM attempt to capture these relationships using earthquake data (Marshall et al., 2023). However, improvements can be achieved by integrating high-resolution relocated catalogs. Besides the major fault planes, minor faults in and around the SFV are illuminated by small magnitude seismicity such as Cluster 1 (Figure 3).

Our focal mechanism solution for the largest event in Cluster 1 ($M_L$ 1.48) suggests strike-slip faulting with a minor thrust component, consistent with the regional tectonic setting (Figure 5a). This demonstrates the potential of dense nodal arrays to resolve small-scale fault structures and kinematics, providing valuable inputs for seismic hazard models. The Cluster 1 fault zone is ~5.3 km by 4 km, and the Hauksson et al. (2012) catalog shows that it has produced $M_w$ 3 events every few years since 1988 (Figure 5d). In 2014, it produced a moderate magnitude $M_b$ 4.4 event, indicating a potential for nucleating larger magnitude events. In our 1-month of recordings, we detected three events along the cluster 1 fault zone compared to the SCEDC catalog annual average of 2.3 events. These events, often overlooked in regional catalogs, can serve as indicators of larger seismic moment releases and offer critical insights into fault interactions, and stress accumulation in urban basins. Temporal and spatial clustering of small events can indicate areas of stress concentration that may eventually culminate in larger ruptures. This phenomenon has been documented in several major earthquake sequences, where retrospective analysis revealed accelerated microseismicity preceding the mainshock (Bouchon et al., 2013; Kato et al., 2012). By improving detection capabilities in urban areas, we can better identify such potential nucleation zones, providing critical inputs for time-dependent seismic hazard models. The spatial distribution of these small magnitude events also helps constrain fault segmentation and interaction zones, where stress transfer between adjacent structures may facilitate or inhibit rupture propagation during larger events.

For example, the Oakridge fault zone (ORFZ) associated with the 1994 Northridge earthquake is a blind thrust fault that extends from the western SFV to the Santa Barbara channel (Jennings et al., 2010; Figure 7a). Geologic interpretation of structural features and fault interactions along the ORFZ, northwest of the SFV has been a subject of controversy with earlier studies before the Northridge earthquake suggesting thin-skinned deformation (Yeats et al., 1988), and post-Northridge cross-section balancing studies indicating thick-skinned deformation structures (Carena & Suppe, 2002; Huftile & Yeats, 1995, 1996). The ORFZ is represented within the 3D SCEC CFM as a simple thrust fault plane (Figure S4), however, cross-section views of relocated seismicity along the fault from Hauksson et al. (2012), highlights southward dipping seismicity along a secondary fault plane that is antithetic to the main thrust (Figure 7b).

Although, a potential proto-ORFZ has been highlighted in geologic studies (Huftile & Yeats, 1995), these faults have not been included in any major databases including the SCEC CFM and California Geological Survey. The antithetic fault branches from the main ORFZ around 10 km depth and has continuity along strike for >20 km (Figure 7b). Multi-fault interactions like this create structural complexity that can potentially lead to larger magnitude earthquakes (Aochi & Kato, 2010). Early aftershocks of the Northridge earthquake reactivated the antithetic fault (Video S1), and additional seismic swarms can be observed in the Hauksson catalog up until 2015 (Hauksson et al., 2012). These faults can host moderate magnitude seismicity, like the $M_b$ 5.1 Northridge aftershock on profile D-D′ (Figure 7b). The focal mechanism for the $M_b$ 5.1 event indicates dominantly thrust faulting with a minor strike slip component, dipping at 32°. Incorporating these antithetic faults into a refined 3D SCEC CFM fault representation can provide improved fault rupture estimates during seismic hazard evaluations. Improved fault geometries would help to better capture the strain partitioning between the main thrust and antithetic faults, providing more accurate rupture propagation barriers and segment boundaries, and account for potential multi-fault rupture scenarios that could cause larger earthquakes that are currently unaccounted for in hazard models (Kyriakopoulos et al., 2017). These refinements are particularly important for dynamic rupture simulations, as complex fault interactions can amplify shaking intensity and duration in areas at risk of seismic damage such as the SFV (Carena & Suppe, 2002).

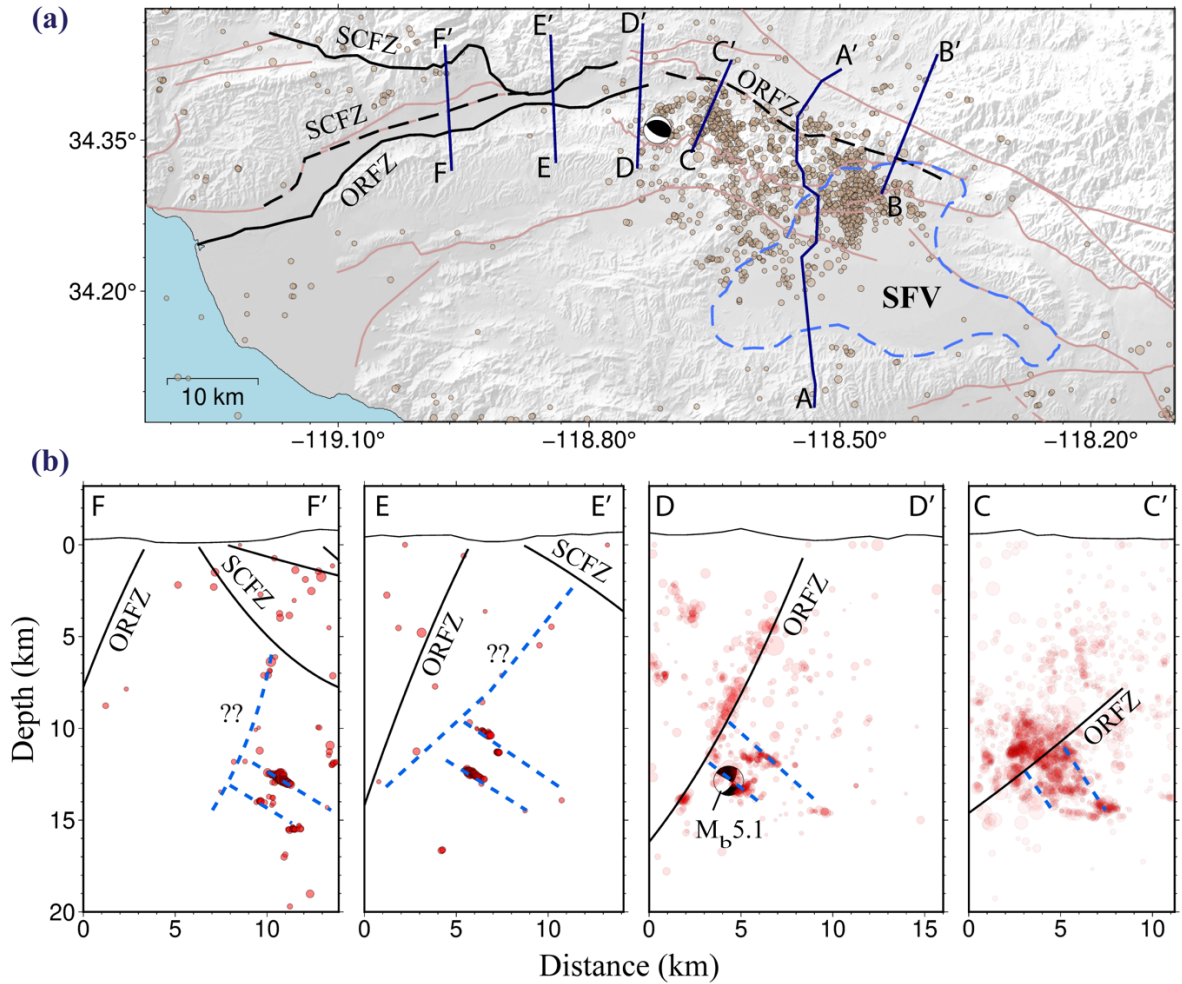

Figure 7. Improving fault geometry estimates with small magnitude seismicity. (a) Topographic relief map showing the Oak Ridge Fault Zone (ORFZ) geometry. Earthquakes ($M_w$>2.5) from Hauksson et al. (2012) are shown as brown circles. Selected cross-sections (dark blue lines) are drawn across the ORFZ. The SFV basin is outlined with the light blue dashed line. The $M_b$ 5.1 focal mechanism is from Yang et al. (2012) (b) Interpreted antithetic faults are shown with dashed blue lines. Projected seismicity and SCEC CFM fault planes are similar to Figure 5. Timelapse view of the seismicity from 1981-2018 along profiles C-F is shown in Video S1.

Finally, the proportion of small to large earthquakes in a catalog is crucial for seismic hazard estimation (Geffers et al., 2022). The central portions of the SFV are devoid of CI network stations (Figure 1). Installing two additional stations, which can be node stations, west and east of Line 1 within the SFV close to the proposed locations in Figure S6 would potentially improve the detectability of the regional network to capture low magnitude seismicity and improve the hypocentral estimates for located events in the SCEDC catalog.

## Conclusion

Our study utilizes a dense nodal array with ML techniques to detect and characterize small magnitude seismicity in the urban environment of the San Fernando Valley (SFV). We effectively addressed the challenges of high anthropogenic noise by omitting uncorrelated phase picks while preserving coherent seismic signals, thus improving computational efficiency for event association. This approach identified 62 events within a month-long period, 36 of which were relocated to obtain improved location estimates. The increased number of event phase picks from our dense array improved our constraints on event locations. Travel-time estimates from our catalog hypocenters matched observed event arrivals better when compared to shared regional catalog events, demonstrating the potential for reducing location uncertainties. We highlight new insights into fault activity and seismic hazard within the SFV and along the Oak Ridge fault zone, where fault interactions should be examined for their potential to promote rupture propagation, increasing the risk of seismic hazard beneath the SFV. Lastly, we evaluated monitoring limitations of the regional network for detecting small magnitude events and proposed infill permanent station locations that can address these issues. These findings highlight how dense nodal deployments can address monitoring gaps in seismically active urban areas, though challenges remain in processing large datasets and differentiating legitimate seismic signals from cultural noise sources.

## Data and Resources

The earthquake catalog and associated data archived at the Southern California Earthquake Data Center (SCEDC) were collected by the Southern California Seismic Network (SCSN), a cooperative project of Caltech and the USGS. The facilities of EarthScope Consortium were used for access to some waveforms, related metadata, and/or derived products used in this study. These services are funded through the National Science Foundation's Seismological Facility for the Advancement of Geoscience (SAGE) Award under Cooperative Agreement EAR-1724509. The published [catalog](#) of relocated seismicity in Southern California from 1981-2018, [focal mechanisms](#) for the relocated events, and the [aftershock catalog](#) of the 1971 San Fernando earthquake were obtained from the SCEDC website special datasets. Three-dimensional fault planes from the SCEC CFM were accessed using the [CFM explorer website](#). Additional faults are from Juárez-Zúñiga and Persaud (2025). Hyperlinks are embedded in the electronic version. Figures were created using the Matplotlib and PyGMT in python (Hunter & Dale, 2007; Uieda et al., 2021). Associated algorithms tested in this study include GaMMA, Pyocto, and REAL (Münchmeyer, 2024; Zhang et al., 2019; Zhu et al., 2022). Supplemental Material for this article includes 6 figures, 1 video, and 3 csv files. A zenodo archive with the preprocessing scripts will be included with the final publication.

## Acknowledgments

We thank the San Fernando Valley residents and business owners for hosting the seismic stations and the 29 volunteers for dedicating their time to install and retrieve the nodal instruments – including Young Ho Aladro, James Atterholt, Ettore Biondi, Alex Clayton, Doug Clayton, Robert

Clayton, Alex Goseyun, Catherine Hanagan, Alan Juarez-Zuniga, Sandra Juarez-Zuniga, Yida Li, Adam Margolis, Joses Omojola, Patricia Persaud , Katleho Ramotso, Samantha Rios, Benjamin Sadler, Ashlyn Schneida, Rajani Shrestha, Elaine Tagge, Valeria Villa, Matthew Wang, Debbie White, Xiaozhuo Wei, Yan Yang, Yifan Yu, and Caifeng Zou. This work was supported by the U.S. Geological Survey award G24AP00067, Statewide California Earthquake Center award 24148, and the National Science Foundation awards 2105320 and 2317154. The SCEC contribution number is 14192. The nodal seismic instruments were provided by the EarthScope Consortium through the EarthScope Primary Instrument Center at New Mexico Tech. Data collected is available through EarthScope. These services are funded through the National Science Foundation's Seismological Facility for the Advancement of Geoscience (SAGE) Award under Cooperative Agreement EAR-1724509.

List of Figure Captions

**Figure 1**. Topographic map of the study area showing the San Fernando Valley nodal array and the Southern California Seismic Network (SCSN) stations used in this study with selected stations labeled. Black lines are faults from the SCEC Community Fault Model (CFM) 6.1 (Marshall et al., 2023). Focal mechanisms of the 1971 San Fernando and 1994 Northridge earthquakes are shown (USGS, 2025). Brown circles are the Mw>2.5 earthquakes from the Hauksson et al. (2012) catalog from 1981 to 2018. Yellow fill in the inset map represents the outline of the major sedimentary basins in Southern California. Major faults and fault zones are the MH-Mission Hills fault, SAF-San Andreas fault, SF-San Fernando fault, SS-Santa Susana fault, ORFZ-Oak Ridge Fault Zone or Northridge thrust, SGFZ-San Gabriel Fault Zone, and SMFZ-Sierra Madre Fault Zone. Additional faults from Juárez-Zúñiga and Persaud (2025) are BCF-Benedict Canyon fault, CF- Chatsworth fault, LF-Leadwell fault, and NLF-North Leadwell fault.

**Figure 2**. Preprocessing of phase picks for event association and comparison of noise levels of the node and broadband stations. (a) Raw picks before preprocessing from October 30, 2023 (UTC) with the x-axis shown in local time. (b) Preprocessed picks after removing isolated detections. Fewer picks speed up the association process. Similar plots for November 05, 2023 (UTC) are shown in Figure S5. (c) Mean noise levels for vertical components of nodal and broadband stations for the full recording period. Relatively quiet broadband instruments are outside the SFV basin, and collocated nodal and broadband instruments within the basin have similar noise levels for periods below 10 s. NHNM: New High Noise Model, NLNM: New Low Noise Model (Peterson, 1993).

**Figure 3**. Map of relocated local events within 10 km of installed stations colored by hypocentral depth. (a) Relocated events are shown as stars. The locations of the A-A', B-B', and C-C' cross-sections in Figure 5 are shown with solid blue lines. Fault names and station are the same as Figure 1. The red circle near node station 371 marks the location of the spurious SCEDC catalog event with waveforms shown in Figure S3. Cluster 1 with three relocated events is highlighted with the brown rectangle. (b) A zoomed-in image of the cluster 1 events. A representative fault plane that aligns with the M> 4.4 aftershock seismicity from Hauksson et al. (2012) is shown with the dashed black line.

**Figure 4**. Comparison of predicted P- and S-wave arrival times for two different event locations for the ML 1.48 event shown in Figure 3. P- and S-wave arrivals in red and blue, respectively, were calculated using a 1D velocity model (Hadley & Kanamori, 1977) to compare event locations from our study (left) and the SCEDC catalog (right) shown in Figure 6b. Waveforms are sorted by epicentral distance. The SCEDC hypocenter gives a poor match and appears later than observed arrivals for stations <10 km from the epicenter. Different start times were used to crop the waveforms to ensure consistency with both catalogs. (inset) The epicentral locations from the SCEDC and this study for the ML 1.48 event shown in Figure 3. Nodal and broadband stations are shown as blue triangles and grey squares, respectively.

**Figure 5**. Cross-sections showing our relocated events with stars colored by hypocentral depth, the background seismicity with red circles (Hauksson et al., 2012), and aftershocks from the 1971 San Fernando earthquake with blue circles (Mori et al., 1995). Profile locations are shown in Figure 3. (a) Projected seismicity within 4 km of profile A-A'. The ORFZ and SFF faults converge beneath the San Fernando Valley. Basin depth and known faults are modified after (Langenheim et al., 2011). The background seismicity shows the interpreted dip of the cluster 1 fault zone. Focal mechanisms for the ML 1.48 and Mb 4.4 events indicate an oblique strike-slip fault, with a dip ranging between 71-72˚, and minor reverse component. (b) Projected background seismicity within 2 km of profile B-B'. Projected fault locations from the SCEC CFM 6.1 are shown as solid lines. Relocated events are linked to seismicity along the SGFZ and ORFZ. (c) Projected seismicity and faults along profile C-C'. Our relocated events are located below the major event clusters. (d) Magnitude versus time distribution of events in the Cluster 1 fault zone from Hauksson et al. (2012) showing quasi-continuous and ongoing activity.

**Figure 6**. Magnitude frequency distribution for our catalog compared to the SCEDC catalog of events around SFV (2010-2023). The blue histogram includes all 36 relocated events detected in this study.

**Figure 7**. Improving fault geometry estimates with small magnitude seismicity. (a) Topographic relief map showing the Oak Ridge Fault Zone (ORFZ) geometry. Earthquakes (Mw>2.5) from Hauksson et al. (2012) are shown as brown circles. Selected cross-sections (dark blue lines) are drawn across the ORFZ. The SFV basin is outlined with the light blue dashed line. The Mb 5.1 focal mechanism is from Yang et al. (2012) (b) Interpreted antithetic faults are shown with dashed blue lines. Projected seismicity and SCEC CFM fault planes are similar to Figure 5. Timelapse view of the seismicity from 1981-2018 along profiles C-F is shown in Video S1.

List of Author Details

Corresponding author: Joses Omojola (jomojo1@arizona.edu)
Second Author: Patricia Persaud (ppersaud@arizona.edu)

**Full mailing address** for both authors (Same building location for both authors)
1040 E 4th St, GS 556, Tucson, AZ 8571938